\documentclass[twocolumn,showpacs,preprintnumbers,superscriptaddress,aps,pra,10pt]{revtex4-1}

\usepackage{amsmath}
\usepackage{amssymb}
\usepackage{amsthm}
\usepackage{bm}
\usepackage{graphicx}

\newcommand{\eq}[1]{Eq.~(\ref{#1})}
\newcommand{\fig}[1]{Fig.~\ref{#1}}
\newcommand{\Tr}{{\rm Tr}}
\newcommand{\E}{{\rm E}}
\newcommand{\mE}{\mathcal{E}}
\newcommand{\mL}{\mathcal{L}}
\newcommand{\cPr}[1]{{\rm Pr}[{#1}]}
\newcommand{\erf}{{\rm erf}}
\newcommand{\Aw}{\langle \hat{A} \rangle _w}
\renewcommand{\Re}{{\rm Re}}
\renewcommand{\Im}{{\rm Im}}

\begin{document}

\title{Statistical hypothesis testing by weak-value amplification: Proposal and evaluation}

\author{Yuki Susa}
\email{susa@th.phys.titech.ac.jp}
\affiliation{Department of Physics, Tokyo Institute of Technology, Tokyo, Japan}
\author{Saki Tanaka}
\affiliation{Graduate School of Science and Technology, Keio University, Yokohama, Kanagawa, Japan}
\date{\today}

\begin{abstract}
We study the detection capability of the weak-value amplification on the basis of the statistical hypothesis testing.
We propose a reasonable testing method in the physical and statistical senses to find that the weak measurement with the large weak value has the advantage to increase the detection power and to reduce the possibility of missing the presence of interaction.
We enhance the physical understanding of the weak value and mathematically establish the significance of the weak-value amplification.
Our present work overcomes the critical dilemma of the weak-value amplification that the larger the amplification is, the smaller the number of data becomes, because the statistical hypothesis testing works even for a small number of data.
This is contrasted with the parameter estimation by the weak-value amplification in the literature which requires a large number of data.
\end{abstract}


\maketitle


\section{INTRODUCTION}
\label{Sec. 1}

The ``weak-value amplification" has been studied as a promising technique for improving an accuracy of a precision
measurement~\cite{Resch2008, Steinberg2010}. The concept of the weak value comes from the weak measurement which was proposed by Aharonov and his co-workers in 1988~\cite{Aharonov1988, Aharonov2005, Aharonov2005_2}.
Originally, the weak measurement was introduced as an example of the two-state-vector formalism for intuitive understanding of the time irreversibility of a measurement in a quantum system~\cite{Aharonov2008}, where the ``weak" means the weak coupling between the two quantum systems: the measured system and the measuring probe.
Usually, we assume that the interaction Hamiltonian is of the von Neumann type which gives a displacement of the order of the coupling constant to the probe distribution~\cite{Jozsa2007}.
An important point of the weak measurement is the postselection of the measured system state after the interaction.
By this operation, the weak value shows up as the shift of the expectation value of the probe position or momentum.
The weak value can be outside the eigenvalue range of the system observable by choosing the postselected state of the system almost orthogonal to its initial state~\cite{Duck1989}.
We call this effect the weak-value amplification (WVA).
If we choose an appropriate posteselected state, the shift of the expectation value of probe position or momentum is enhanced larger than the one given by the measurement without postselection which coincides with the coupling constant.
We parenthetically note that for a large coupling constant, the measurement without postselection is sometimes called the strong measurement.
Therefore, we may hope to extract the information about the coupling constant even if the constant is smaller than a noise. 

Reference~\cite{Dressel2014} introduced the basic concept of the weak value and its application.
It is known that the amplified shift has an upper bound if we take Gaussian for the initial probe state~\cite{Wu2011}.
Especially, when the measured system is a two-state system, some researchers have shown the upper bound analytically without any approximation~\cite{Zhu2011, Koike2011, Nakamura2012}.
The initial measuring probe wave function which maximizes the amplification factor was studied in Refs.~\cite{Susa2012, Lorenzo2013, Susa2013}.
The WVA was confirmed in several experimental studies.
Setting two polarizers in front and behind of a birefringent crystal, the weak value was measured and found to become large by arranging two polarizers almost orthogonal~\cite{Ritchie1991}.
Hosten and Kwiat observed the spin Hall effect of light with the WVA~\cite{Hosten2008}.
Dixon {\it et al}. monitored the laser deflection by the small tilting Piezo mirror in the Sagnac interferometer and measured the angle of the mirror in precision~\cite{Dixon2009}.
Viza {\it et al}. demonstrated the velocity measurement of the longitudinal moving mirror in the Michelson interferometer~\cite{Viza2013}.

There are theoretical researches about technical utilities of the WVA.
Nishizawa {\it et al}. compared the signal and the shot noise in an optical interferometer~\cite{Nishizawa2012}.
Jordan {\it et al}. showed that the WVA has the error tolerance of the systematic error~\cite{Jordan2014}.
Lee and Tsutsui discussed the causes of the errors in the weak measurement with finite data and the merit of the WVA~\cite{Lee2014}.
However, the problem is that the larger the amplification factor is, the smaller the success probability of the postselection. The small number of detectable events leads to the possible disadvantage of the interaction parameter estimation using the WVA as argued in~\cite{Knee2013, Tanaka2013, Ferrie2014, Knee2014, Knee2014_2},
while some researchers have mentioned that the data loss by postselection usually need not to be considered in practical cases~\cite{Zhu2011, Vaidman2014}.

Lately, the interaction detection capability of the WVA has been focused on~\cite{Lee2014, Ferrie2014}.
In this problem, using the WVA, we want to decide whether the interaction exists or not in an indirect quantum measurement process.
The result of Ref.~\cite{Hosten2008} suggests that the detection capability of the WVA is experimentally utilitarian.
To theoretically study such a problem, the statistical hypothesis testing is a well-known method~\cite{Rao1973, Kiefer1987}.
The hypothesis testing does not require a large number of data like the estimation, and the accuracy of testing is normally independent of the number of data.
Therefore, we note that the hypothesis testing works well for a small number of data given by the weak measurement.
In Ref.~\cite{Ferrie2014}, however, the authors claim that the WVA is suboptimal for the interaction detection.
Their conclusion is based on the discussion of the likelihood-ratio test with some debate~\cite{Vaidman2014, Kedem2014,Ferrie2014a}.
It seems to us, however, we have to pay more attention to the well-known fact that the likelihood-ratio test is not appropriate to the detection of the interaction.
More precisely, because the interaction detection problem is a two-side test, a uniformly most powerful unbiased (UMPU) test is the standard procedure to solve this problem.

In this paper, we propose a statistical hypothesis testing based on the physical intuition, and analytically evaluate the interaction detection capability of the WVA.
To determine whether the interaction exists or not, we give a decision function such as the measurement outcome divided by the initial fluctuation of the measuring probe distribution [$\bigl|x\bigr|/\sigma$ below \eq{eq:SNdecision}].
Under this decision function and the particular condition for the weak value, we find that the WVA can supersede the ordinary measurement without postselection.
We can say a large weak value increases the detection power.
More precisely, the advantage of the WVA is the reducing of the possibility of missing the presence of the interaction with the false alarm rate fixed.
Our result is suggestive for the interpretation of the weak value.
Some researchers simply take the physical intuition of the WVA for granted with the approximation in which the weak value amplifies the shift of the probe wave function.
In this work, we show without any approximation that the weak value itself determines the superiority or inferiority of each measurement, the weak measurement and the ordinary measurement, in the detection capability.
We emphasize that our result mathematically clarifies the significance of the WVA which justifies the physical intuition.
Throughout this paper, we assume that the initial probe wave function is Gaussian and the measured system is a two-state system as considered in Refs.~\cite{Zhu2011, Koike2011, Nakamura2012}.

Our paper is organized as follows.
In Sec. \ref{Sec. 2}, we give brief reviews of the weak measurement and the statistical hypothesis testing which includes a UMPU test for the two-side test.
Section \ref{Sec. 3} gives the main result of this work.
We propose a proper decision function for the interaction detection and evaluate the statistical errors so-called ``type-1 error" and ``type-2 error."
We show the detection power superiority superiority of the WVA and the required condition for the weak value with analytical derivation.
In Sec. \ref{Sec. 4}, we consider an additive white Gaussian noise model.
In this situation, the result of Sec. \ref{Sec. 3} holds when the unknown extra fluctuation exists. 
We have summary and discussion in Sec. \ref{Sec. 5}.
There, we discuss the case that the data would be unobtainable by failure of the postselection.
Some complicated calculations and the supplementary discussion of the unobtainable case are shown in Appendixes.
We use the unit $\hbar=1$ and the subscripts that ``ps'' means the case of the measurement with the postselection, i.e., the weak measurement and ``nps'' indicates the case of the ordinary measurement, i.e., the measurement with no postselection.


\section{REVIEW OF WEAK MEASUREMENT AND HYPOTHESIS TESTING}
\label{Sec. 2}


\subsection{Weak measurement and probability distributions}
\label{Sec. 2A}

We recapitulate the standard weak measurement process~\cite{Aharonov1988, Jozsa2007, Steinberg2010, Dressel2014} and derive the probability distribution after the measurement.
The weak measurement is described as the indirect quantum measurement formalism.
To carry out the weak measurement, initially we prepare a measured system $\mathcal{H}$ and a measuring probe $\mathcal{K}$.
(Hereafter, we omit the indexes $\mathcal{H}$ and $\mathcal{K}$ unless otherwise stated.)
The initial state of the measured system $\mathcal{H}$ is a preselected state $|i\rangle_\mathcal{H}$
and the initial state of the probe system $\mathcal{K}$ is $|\psi \rangle_\mathcal{K} = \int \psi (x) | x \rangle_\mathcal{K} dx $, which is taken as the Gaussian profile;
\begin{equation}
\label{eq:wfGauss}
\psi (x) = C e^{- \frac{x^2}{4 \sigma^2} } 
,~~C ^2 =\frac{1}{\sqrt{2\pi \sigma^2} } ,
\end{equation}
where the $x$ represents the position of the probe.
We denote $\hat{\rho}_i:=|\psi \rangle \langle \psi|$ for the probe state in a density matrix expression for later convenience.
We assume the von Neumann interaction so that the time evolution operator given by 
$\hat{U}=\exp(-i g  \hat{A} ^\mathcal{H} \otimes \hat{p} ^\mathcal{K} )$, 
where $\hat{A} ^\mathcal{H}$ is an observable defined in $\mathcal{H}$ and $\hat{p} ^\mathcal{K}$ is a momentum operator satisfying $[\hat{x} ^\mathcal{K} ,\hat{p}^\mathcal{K}] =i$ defined in $\mathcal{K}$. The parameter $g$ indicates an unknown coupling strength.
The interaction produces the state of the combined system as
\begin{equation}
\label{eq:interactedstate}
    \hat{\rho}_\mathrm{int}^{\mathcal{H}\otimes \mathcal{K}}
       :=\hat{U} (|i\rangle_\mathcal{H}\langle i| 
              \otimes \hat{\rho}_{\rm i}^\mathcal{K}) \hat{U}^\dagger. 
\end{equation}
Finally, we postselect the measured system state $|f\rangle_\mathcal{H}$. 

Here, we assume that the interaction strength $g$ is sufficiently small that the first approximation in $g$ is available.
The postselection translates the wave function by $g\Re \Aw$, i.e.,
\begin{equation}
\label{eq:translate}
\psi ( x ) \rightarrow \psi ( x-g\Re \Aw),
\end{equation}
where the $\Aw$ is called the weak value defined by
\begin{align}
\Aw:= \frac{\langle f| \hat{A} |  i \rangle}{\langle f| i\rangle}.
\end{align}
We can easily see that the weak value has a generally complex value, and becomes infinitely large when the postselected state $\langle f|$  is almost orthogonal to the initial state $| i\rangle$.
Because of the large $\Aw$, we find that the shift of the probe wave function gets large as can be seen from \eq{eq:translate}.
This implies that the coupling constant $g$ is effectively amplified.
This effect is called ``weak-value amplification" (WVA)~\cite{Hosten2008}.
We emphasize that the postselection is essential in the WVA.

For the later discussion, we carry out full order calculation of the final probability distribution of the measuring probe concentrating on the two-state system case for the measured system.
The final probe state becomes
\begin{equation}
\label{eq:psstate}
   \hat{\rho}^\mathcal{K}_\mathrm{ps} 
    := \frac{ \Tr_{\mathcal{H}}\big[ 
            ( |f\rangle_\mathcal{H}\langle f|  \otimes \hat{I}^\mathcal{K}) 
                \hat{\rho}_\mathrm{int}^{\mathcal{H} \otimes \mathcal{K}} ]}
              {\Tr\big[ 
            (|f\rangle_\mathcal{H}\langle f|  \otimes \hat{I}^\mathcal{K}) 
              \hat{\rho}_{\rm int}^{ \mathcal{H} \otimes \mathcal{K} } \big]}.
\end{equation}
Here we note that the denominator coincides with the success probability of the postselection of the $|f \rangle$.
We obtain the large amplification when the probability becomes small.
However, the small probability reduces the accuracy of the estimation~\cite{Knee2013, Tanaka2013, Ferrie2014}.

From the state (\ref{eq:psstate}), we can calculate the final distribution which can be tested in a real experiment.
Following the standard discussions~\cite{Aharonov1988, Aharonov2005, Ritchie1991,Hosten2008}, we study the probe distribution in the position basis.
It is straightforward to have the position probability distribution of the final probe as 
\begin{align}
\label{eq:fps}
& f_{\mathrm{ps}}(x|g)
=\Tr \left[\hat{\rho}^\mathcal{K}_\mathrm{ps}| x \rangle_{\mathcal K} \langle x | \right] 
=\frac{| \langle x|_\mathcal{K}  \langle f | \hat{U} | i \rangle _\mathcal{H}  | \psi \rangle _\mathcal{K} | ^2}{
| \langle f | \hat{U} | i \rangle _\mathcal{H}  | \psi \rangle _\mathcal{K}| ^2 }  \notag\\
&=\frac{1}{2\sqrt{2\pi\sigma ^2} }
\frac{1}{1 + | \Aw |^2 + (1 - | \Aw |^2 )  e^{-\frac{g^2}{2\sigma ^2} }} 
 \notag \\&  \times
\left\{ 
 \begin{array}{l}
( 1 + | \Aw |^2 +2 \Re \Aw )  ~e ^{-\frac{(x-g)^2}{2 \sigma^2}} \\
+( 1 + | \Aw |^2 -2 \Re \Aw)  ~e ^{-\frac{(x+g)^2}{2 \sigma^2}} \\
+2 ( 1 - | \Aw |^2 )   ~e ^{-\frac{x^2+g^2}{2 \sigma^2}} 
 \end{array} 
 \right\} ,
\end{align}
where we have used the assumption that the measured system is a two-state system, i.e., $\hat{A}^{2}=1$.
Here, we define that $|+\rangle$ and $|-\rangle$ are the eigenstates of $\hat{A}$ in the measured system, and our discussion holds without loss of generality even if the observable is set as $\hat{A}=|+\rangle\langle+|-|-\rangle\langle-|$.

We also give the position distribution observed in the measurement without postselection to be compared with the distribution (\ref{eq:fps}).
The final probe state is given by taking the partial trace of the combined state (\ref{eq:interactedstate}) as
\begin{align}
 \hat{\rho}_\mathrm{int}^{\mathcal{K}}
= \Tr _\mathcal{H} \bigl[  \hat{U}  ( | i \rangle \langle i |   \otimes | \psi \rangle \langle \psi | ) \hat{U} ^\dagger \bigr].
\end{align}
This density matrix gives the position probability distribution as 
\begin{align}
\label{eq:fnps}
& f_{\mathrm{nps}} (x|g) 
=\Tr \left[  \hat{\rho}_\mathrm{int}^{\mathcal{K}} | x \rangle _\mathcal{K}  \langle x | \right ]
= | \langle x | \hat{U} | \psi \rangle _\mathcal{K}  | i \rangle _\mathcal{H} | ^2 \notag \\
& = \frac{1}{\sqrt{2\pi\sigma^2} } 
\left\{
|\langle + | i \rangle |^2 ~e ^{- \frac{(x-g)^2}{2 \sigma ^2} }
+|\langle - | i \rangle |^2 ~e ^{- \frac{(x+g)^2}{2\sigma ^2} }
\right\} .
\end{align}

\begin{figure}[tbp]
\centering
\includegraphics[width=8.6cm]{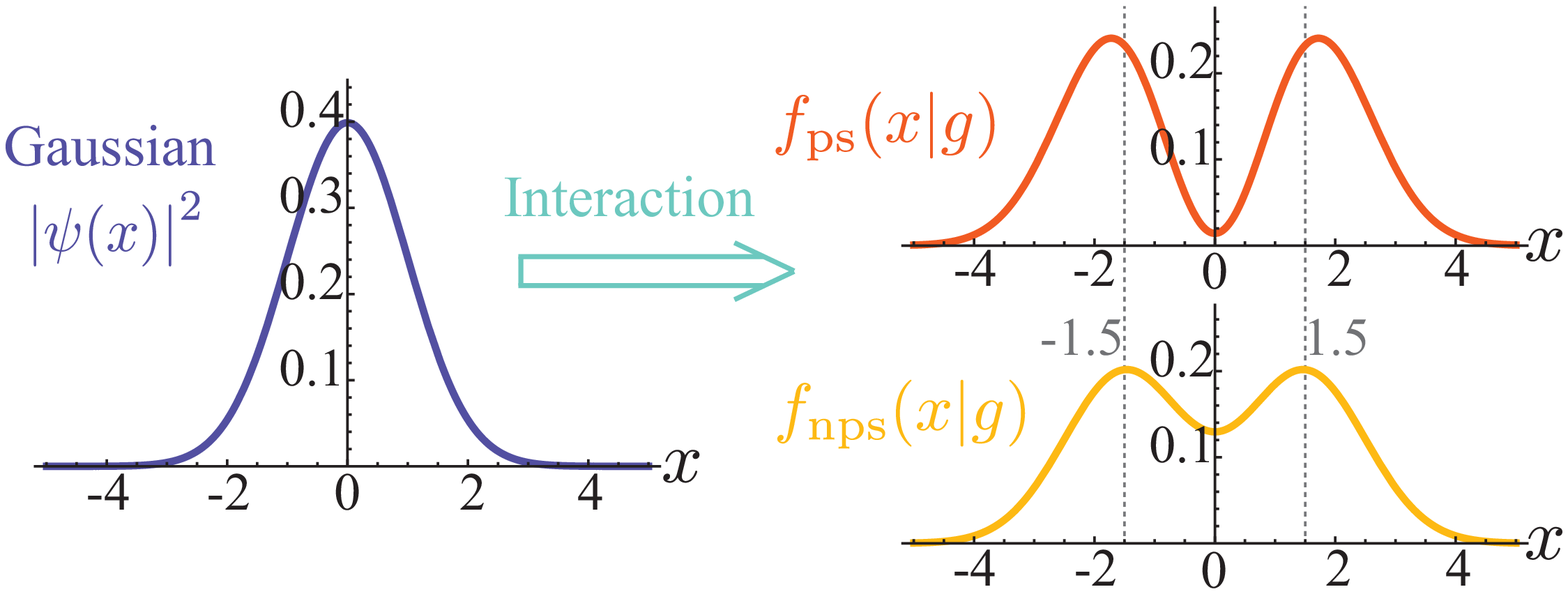}
\caption{
(Color online) A sketch of the distribution functions when $\sigma =1$ and $g=1.5 $. 
The blue (left) curve is the probe distribution before the interaction.
The red (upper right) and the yellow (lower right) curves are the ones after the interaction with postselection and without postselection, respectively.
We choose the initial state of the measured system as $| i  \rangle = (| + \rangle + | - \rangle) /\sqrt{2}$, and the postselected state which satisfies the weak value $\Re \Aw =0$ and $| \Aw | =5$.}
\label{fig1}
\end{figure}
Figure \ref{fig1} shows the position probability distribution transition from a Gaussian probe as the initial probe to the final probes after the interaction.
We can see that the postselection causes the difference from $f_{\mathrm{nps}}$ to $f_{\mathrm{ps}}$.
The weak value $\Aw$ plays an important role in determining the distribution change.
Note that the condition $\Re \Aw=0$ makes the two peaks at the same height and that the increases of the imaginary part $\Im \Aw$ depresses the dip deeper while increasing the distance of the two peaks.
This may not be very obvious, but is justified by the closer algebraic examination of the expression (\ref{eq:fps}).
With an appropriate postselection, we can obtain peak shifts larger than the ordinary peak-to-peak distance~\cite{Duck1989}, which is given by the coupling constant times the eigenvalues of the observable as indicated in \fig{fig1}.
Note that in the case of $\Aw=\pm1$, the postselection reduces to the projective measurement of $\hat{A}$.
A similar discussion goes through for $|i\rangle=|+\rangle$ or $|-\rangle$ in the case without postselection.


\subsection{Hypothesis testing for detection problem}
\label{Sec. 2B}

Generally, the statistical hypothesis testing is used for a mathematical decision from the data of measurement~\cite{Ferguson1967, Rao1973, Kiefer1987}.
Our aim of this paper is the evaluation of the interaction detection capability by the WVA in which the size of data is not necessarily large.
In the classical theory of statistics, the hypothesis testing provides a better solution for such evaluation than the estimation theory as explained below.

The estimation accuracy is usually evaluated by the Fisher information.
More precisely, the mean squared error of the estimator approaches the inverse of the Fisher information for an infinite number of data.
Then, the Fisher information gives a good description for the accuracy of the estimation with a large number of data~\cite{Roy2004, Johann1994, Nagaoka2005, Fujiwara2006}.
In the weak measurement, the failed postselection makes data loss, so that the estimation theory is not a suitable method for analyzing the relative power of the WVA when the number of experiments is limited.
On the other hand, the hypothesis testing can be evaluated by the error probability regardless of the number of data.
So, our discussion of the WVA based on the hypothesis testing theory works even when we have a small number of data.

The hypothesis testing is a statistical inference to decide which hypothesis is appropriate from the measurement results in the two contradictory hypotheses, the null and the alternative. 
For the detection problem, we take the hypotheses as follows: (a) the null hypothesis $H_0$: absence of the interaction (i.e., $g=0$), (b) the alternative hypothesis $H_1$: presence of the interaction (i.e., $g \neq 0$).
This hypothesis testing problem is a two-side test in the sense that the estimated parameter deviates in either direction ($g>0$, $g<0$) from the null hypothesis ($g=0$).
We derive a decision function which takes the binary value as $d(x)=0$ or $1$.
If it is $0$, we reject the alternative hypothesis, and if it is $1$, we reject the null one.
A range of measurement results $x$ such that $d(x)=0$ is called the acceptance region. 
When the measurement result falls within this region, the alternative hypothesis is rejected. 
A set of the outside of the acceptance region is called the rejection region.
When the measurement result is within this region, the null hypothesis is rejected.
Since the decision function $d(x)$ is independent of the coupling parameter $g$, sometimes we get the wrong indications from the decision function.

Such wrong indications are classified to the two types,
(i) the type-1 error: even if the null hypothesis is true, we wrongly reject it, (ii) the type-2 error: even if the null hypothesis is false, we wrongly accept it as truth.
In our case, the type-1 error represents ``there is no interaction but we wrongly guess the interaction exists, i.e., falsely alarmed."
The type-2 error means ``the interaction indeed exists, but we wrongly guess the interaction does not exist, i.e., miss the presence of the interaction."
The probability of the type-1 error is calculated as the integration of the probability distribution function over the rejection region of the null hypothesis when the null hypothesis is true.
The probability of the type-2 error is calculated as the integration of the probability distribution function over the acceptance region of the null hypothesis when the alternative hypothesis is true.
The smaller probabilities of these errors become, the better the testing is.
Since it is difficult to make the two types of error small simultaneously,
we make the probability of the type-2 error as small as possible while suppressing the probability of the type-1 error under a certain significance level as the general strategy~\cite{Rao1973}.
In this paper, we compare the weak measurement and the measurement without postselection by evaluating the probabilities of the two types of error for each measurement.

To carry out the test with the small probabilities of the errors, an adequate decision function is necessary.
The uniformly most powerful (UMP) test is one of the good testings in the general strategy~\cite{Kiefer1987}. 
The Neyman-Pearson lemma is a famous example for explaining the UMP test~\cite{Nyeman1933}.
This lemma claims the likelihood-ratio test is UMP only when the both hypotheses are simple, 
i.e., $H_0:~ \theta =\theta _0$ and $H_1:~ \theta = \theta _1 $.
The likelihood-ratio test is available even when either or both of the hypotheses is composite, for instance our hypotheses, by using the maximum likelihood-estimator (MLE), e.g., see Ferrie and Combes~\cite{Ferrie2014}.
However, from the statistical inference~\cite{Roy2004, Rao1973}, it is widely known that the MLE without the large number of data is not helpful for extracting the information of a physical system.
In addition to this problem, the UMP test does not exist in the two-side test.
Therefore, the likelihood-ratio test is secondary.

To find an adequate decision function in the two-side test, we introduce the concept of the unbiased test $d(x)$ defined by
\begin{align}
\beta(\theta):= \int d(x) f(x | \theta) dx \ge \alpha,
\end{align}
where the $\alpha$ is a significance level. 
The statistical power $\beta(\theta )$ with $\forall \theta \in \Theta _1$ represents the capability of the detection, i.e., the larger power means the higher detectability of the interaction.
Here $\Theta _1$ is a set of parameters delineating the alternative hypothesis $H_1$.
We can calculate the probabilities of each error as follows: one of the type-1 error is $\cPr{\mE_{1}}:=\beta(\theta_0)$ and one of the type-2 error is $\cPr{\mE_{2}}:=1-\beta(\theta\in\Theta_1)$.
According to the following lemma, we can obtain the UMPU test which is a good test for composite hypotheses such as a two-side test~\cite{Ferguson1967}. 

{\it Lemma.} If the hypotheses are given by the two-side test: $H_0:~\theta=\theta_0$ and $H_1:~\theta\neq\theta_0$ , we assume that the decision function $d(x)$ satisfies
\begin{gather}
\label{eq:UMPUconditionsmooth}
\partial_{\theta} \beta(\theta)= \int d(x) \partial_{\theta} f(x|\theta) dx , \\
\label{eq:UMPUlemmacondition}
\beta(\theta_{0})= \alpha, \\
\label{eq:UMPUlemmacondition2}
\partial_{\theta} \beta(\theta_{0})=0 .
\end{gather}
For an any fixed $\theta _{1}$ such that $\theta _{1}\neq \theta _{0}$, 
if the test $d(x) $ is given by
\begin{align}
d (x) &=\left\{
\begin{array}{l}
\label{eq:UMPUstatistics}
0 ~~~~\text{if} ~ \mathcal{F} (x)<0,\\
r ~~~~\text{if} ~ \mathcal{F} (x)=0,\\
1 ~~~~\text{if} ~ \mathcal{F} (x)>0,
\end{array}
\right.
\end{align}
where
\begin{align}
\label{eq:extendlikehoodfunction}
\mathcal{F}(x) &:=f(x|\theta_{1})-c_{1}f (x|\theta _{0})-c_{2} \partial_{\theta} f (x|\theta)\big| _{\theta=\theta _{0}}
\end{align}
with certain parameters $c_{1}$ and $c_{2}$, the $d(x) $ is the UMPU test.
The $d(x)$ becomes the randomized test $d(x)=r$, and $r~ (0\le r \le 1) $ is the probability to accept the null hypothesis~\cite{randomized test}.

This lemma works for a small number of samples such as the data given by the weak measurement.
In the following section, we propose a test which has a physical meaning, and we check that the test is UMPU on the basis of the above lemma.


\section{HYPOTHESIS TESTING WITH WEAK-VALUE AMPLIFICATION}
\label{Sec. 3}


\subsection{Merit of WVA in interaction detection}
\label{Sec. 3A}

In what follows, we derive the UMPU test for the detection of the presence of the weak interaction $g$ in the two cases.
As mentioned in Sec. \ref{Sec. 2B}, our proposed test is the best one among all unbiased tests.
Then, we compare the best test in the WVA with that in the measurement without postselection by explicit forms.
We evaluate the testing capability, comparing the probabilities for the type-1 and the type-2 errors for the two cases: the weak measurement and the measurement without postselection.
We remark that, in this section, we treat the case that the data are not empty even if there is data loss caused by postselection, i.e., the transition probability $\bigl|\langle f| i \rangle\bigr|^2$ is not zero.
The unobtainable case will be discussed separately in Sec. \ref{Sec. 5}. 

For a fair comparison of the WVA and the measurement without postselection, we establish the UMPU test for each measurement on the basis of the Lemma.
The first step is proposing a suitable decision function as a candidate of the UMPU test.
The decision function must be independent of the unknown parameter $g$.
We have assumed that the initial probe distribution is Gaussian (\ref{eq:wfGauss}) with its variance $\sigma^2$.
Roughly speaking, if there is no interaction, almost all the measurement results will be inside of the initial fluctuation $|x|<\sigma$ and the probability of $|x|>\sigma$ is relatively small.
On the other hand, if the interaction exists, we ought to get some measurement outcome which is deviated from the initial fluctuation, and the probability of $|x|>\sigma$ would become significantly larger.
Precisely, we propose the following decision function:
\begin{equation} 
\label{eq:SNdecision}
d(x) =
\left\{
\begin{array}{l}
0 ~~~~\text{if} ~ |x|/ \sigma  < c,  \\
r ~~~~\text{if} ~ |x|/ \sigma  = c,  \\
1 ~~~~\text{if} ~ |x|/ \sigma  > c,
\end{array}
\right.
\end{equation}
where a critical point $c$ is a positive constant that we can choose as we like.
The rejection region is fixed to $|x|> c\sigma$.
We verify that this decision function (\ref{eq:SNdecision}) is the UMPU test in Sec. \ref{Sec. 3B}.
Additionally, the distribution function after the measurement with $\Re \Aw=0$ or $\bigl| \langle + | i \rangle \bigr|^{2}=\bigl| \langle - | i \rangle \bigr| ^{2}$ becomes an even function in the case of the weak measurement or the measurement without postselection, respectively.
Under these particular situations, we can practically interpret that the testing of the detection problem becomes a one-side test, i.e., $H_0:~ g=0$ and $H_1:~ g>0$.
There is a theorem which gives the UMP test for such a one-side test, and we can show our decision function (\ref{eq:SNdecision}) gives a UMP test (see Sec.~\ref{Sec. 3C} for detail.).

Here, we compare the probabilities of the type-1 and -2 errors obtained by Eqs. (\ref{eq:fps}) and (\ref{eq:fnps}).
First, we consider the type-1 error when the coupling constant is $g=0$ and the measurement result is $|x|>c\sigma$.
Since the distribution functions $f_{\mathrm{ps}}(x|g)$ and $f_{\mathrm{nps}}(x|g)$ coincide at $g=0$,
\begin{align}
f _{\mathrm{ps}} (x | g =0 ) = f_{\mathrm{nps}} (x | g =0 ) 
= e^{-\frac{x^2}{2\sigma ^2}}/\sqrt{2 \pi  \sigma ^2},
\end{align}
the probabilities of the type-1 error of $f _{\mathrm{ps}}(x|g) $ and of $f _{\mathrm{nps}}(x|g) $ are the same as
\begin{align}
\label{eq:1sterror}
\cPr{\mE_{1}}=\beta(0)=1-\erf[c/\sqrt{2}],
\end{align}
where $\erf[x] :=\frac{2}{\sqrt{\pi}} \int _0  ^x e ^{-t^2}dt$ is the error function.
The $\cPr{\mE_{1}}$ can be any significance level by choosing $c$.
Hence, this test suits the standard strategy~\cite{Roy2004}.

\begin{figure}[tbp]
\centering 
\includegraphics[width=8.6cm]{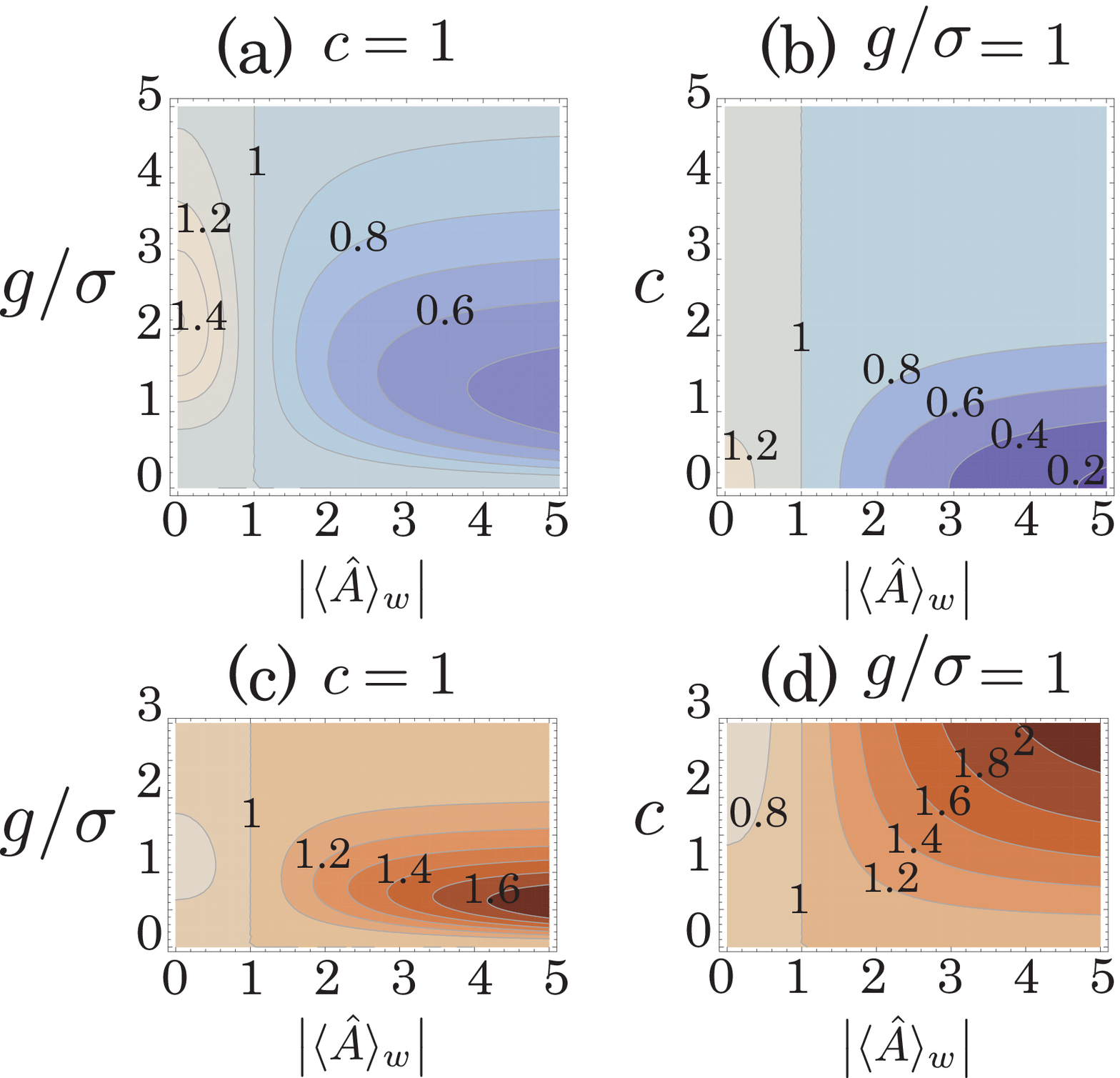}
\caption{(Color online) The contour plots (a), (b) for $\cPr{\mE_{2,\mathrm{ps}}}/\cPr{\mE_{2,\mathrm{nps}}}$ and (c), (d) for $\beta_{\mathrm{ps}}(g)/\beta_{\mathrm{nps}}(g)$ in which the horizontal axis indicates the absolute value of the weak value $| \Aw |$.
The left- and right- sides graphs have the vertical axis indicating the coupling constant divided by the initial fluctuation $g/\sigma$ and the critical point $c$, respectively.
In (a) and (b), the darker blue indicates the smaller value.
In (c) and (d), the darker red indicates the larger value.
}
\label{fig2}
\end{figure} 

\begin{figure*}[tbp]
\centering 
\includegraphics[width=\textwidth]{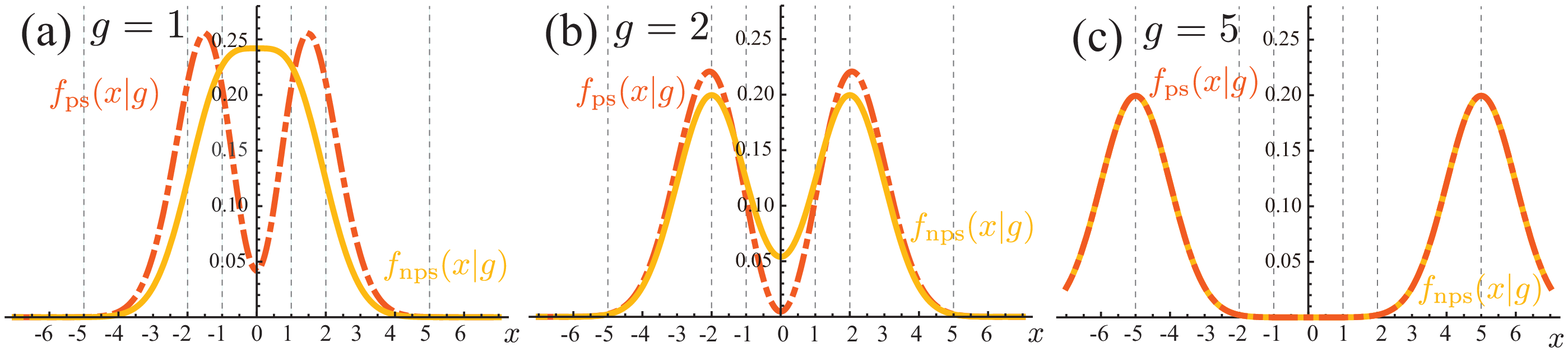}
\caption{(Color online) Plots of the wave functions $f_{\mathrm{ps}}(x|g)$ (the red dashed curve) and $f_{\mathrm{nps}}(x|g)$ (the orange solid curve) against $x$ in three coupling constant cases: (a) $g=1$, (b) $g=2$, and (c) $g=5$.
In the case (c), two plots are almost overlapped.
In these graphs, the other parameters are fixed as follows: $\Re \Aw =0$, $| \Aw |=5 $, $|\langle + | i \rangle |^2 =|\langle - | i \rangle |^2 =1 /2$, and $\sigma=1$.
}
\label{fig3}
\end{figure*}

Next, we evaluate the probabilities of the type-2 error when $g\neq0$ and $|x|<c\sigma$.
In this case, the distribution functions (\ref{eq:fps}) and (\ref{eq:fnps}) are different.
The probability of the type-2 error in the weak measurement is
\begin{align}
\label{eq:error2ps}
&\cPr{\mE_{2,\mathrm{ps}}}=1-\beta_{\mathrm{ps}}(g) \notag \\
&=\frac{1}{2[ 1+ | \Aw |^2
+ ( 1- | \Aw | ^2) e^{-\frac{g^2}{2 \sigma ^2}}]} 
 \notag  \\
&\times 
\left[
\begin{array}{l}  
( 1+ | \Aw| ^2)
\big\{ 
   \erf \big[\frac{c\sigma-g}{\sqrt{2 \sigma^2}} \big] + 
   \erf \big[\frac{c\sigma+g}{\sqrt{2\sigma^2}} \big]   
 \big\}\\
   + 2 ( 1- | \Aw | ^2) ~e^{-\frac{g^2}{2 \sigma ^2}}~ \erf \big[\frac{c}{\sqrt{2}}\big]  
\end{array}
    \right].
\end{align}
In the measurement without postselection, the probability of the type-2 error is given by
\begin{align}
\label{eq:error2nps}
\cPr{\mE_{2,\mathrm{nps}}}&=1-\beta_{\mathrm{nps}}(g) \notag \\
&=\frac{1}{2}\left(
\erf \left[\frac{c\sigma-g}{\sqrt{2\sigma^2}} \right] + 
\erf \left[\frac{c\sigma+g}{\sqrt{2\sigma^2}} \right]\right).
\end{align}
To determine which measurement gives more benefit, it is enough to compare the probabilities (\ref{eq:error2ps}) and (\ref{eq:error2nps}).
We can derive the following equation by arranging the ratio $\cPr{ \mE_{2,\mathrm{ps}}}/\cPr{\mE_{2,\mathrm{nps}}}$, which is symmetric under the sign change $g\leftrightarrow -g$, as
\begin{align}
\label{eq:E2ratio}
&\cPr{ \mE_{2,\mathrm{ps}}}/\cPr{\mE_{2,\mathrm{nps}}}-1 \notag \\
&=\frac{ ( 1- | \Aw | ^2)\left(\frac{2 \erf \left[ \frac{c}{\sqrt{2}} \right]  }{\erf \left[\frac{c\sigma-g}{\sqrt{2\sigma^2}} \right] + \erf \left[\frac{c\sigma+g}{\sqrt{2\sigma^2}} \right] }-1 \right)  e^{-\frac{g^2}{2 \sigma ^2}}   }{ 1+ | \Aw |^2+ ( 1- | \Aw | ^2) e^{-\frac{g^2}{2 \sigma ^2}}  }.
\end{align}
We can see that the inequality
\begin{align}
\label{eq:ineq}
&\cPr{ \mE_{2,\mathrm{ps}}}\leq\cPr{\mE_{2,\mathrm{nps}}} \\
\label{eq:ineq_power}
\Leftrightarrow &\beta _{\mathrm{ps}}(g) \ge \beta _{\mathrm{nps}}(g)
\end{align}
hold for such a weak value that $| \Aw |\geq1$ as shown in Appendix \ref{App. A}.
Therefore, the probability of the type-2 error with the postselection can be less than the one without postselection in a certain case.

We note that the derivative function of the probabilities of the type-2 ratio $\cPr{\mE_{2,\mathrm{ps}}}/\cPr{\mE_{2,\mathrm{nps}}}$ is 
\begin{align}
&\frac{\partial \cPr{\mE_{2,\mathrm{ps}}}/\cPr{\mE_{2,\mathrm{nps}}}}{\partial | \Aw | ^{2}} \notag \\
&=\frac{-2 \left(\frac{2 \erf \left[ \frac{c}{\sqrt{2}} \right]  }{\erf \left[\frac{c\sigma-g}{\sqrt{2\sigma^2}} \right] + \erf \left[\frac{c\sigma+g}{\sqrt{2\sigma^2}} \right] }-1 \right)  e^{-\frac{g^2}{2 \sigma ^2}} 
}
{\big[  1+ | \Aw |^2+ ( 1- | \Aw | ^2) e^{-\frac{g^2}{2 \sigma ^2}}\big]^{2}
}
\le 0.
\end{align}
Then, we find that the ratio $\cPr{\mE_{2,\mathrm{ps}}}/\cPr{\mE_{2,\mathrm{nps}}}$ is a monotonically decreasing function with respect to $| \Aw |^2$.
Similarly, we can show that the detection power ratio $\beta _{\mathrm{ps}}(g) / \beta _{\mathrm{nps}}(g)$ is a monotonically increasing function with respect to  $| \Aw |^2$.

Figure \ref{fig2} shows the ratio $\cPr{\mE_{2,\mathrm{ps}}}/\cPr{\mE_{2,\mathrm{nps}}}$ and $\beta _{\mathrm{ps}}(g) /\beta _{\mathrm{nps}}(g)$ for the three parameters $|\Aw|$, $g/\sigma$, and $c$.
We can see from these graphs that the inequalities (\ref{eq:ineq}) and (\ref{eq:ineq_power}) hold when the weak value $| \Aw |$ is larger than 1.
They also indicate that the WVA well works when the coupling constant divided by the initial fluctuation $g/\sigma$ and the critical point $c$ are relatively small.
We believe that this is the heart of the weak-value amplification which is mathematically well grounded.
On the other hand, if the $g/\sigma$ is large, the large weak value is as helpful to detect as the ordinary measurement.
This property comes from the difference of the distribution functions $f_{\mathrm{ps}}$ and $f_{\mathrm{nps}}$.
The difference can be explained in \fig{fig3}.
When the $g$ is small as shown in \fig{fig3} (a), there is a big difference between $f_{\mathrm{ps}}$ and $f_{\mathrm{nps}}$ for the small $|x|$.
Note that there is no difference between $f_{\mathrm{ps}}$ and $f_{\mathrm{nps}}$ for a large $g$ as we can see from Figs. \ref{fig3} (b) and (c).
The probability of the type-2 error is given by the integration over the interval $[-c\sigma, c\sigma]$.
Since $f_{\mathrm{ps}}$ is smaller than $f_{\mathrm{nps}}$ in the central region of $x$ for a fixed small $g$, the ratio $\cPr{\mE_{2,\mathrm{ps}}}/\cPr{\mE_{2,\mathrm{nps}}}$ becomes small, if we properly choose the critical point $c$.
In contrast, we can see from Eqs. (\ref{eq:error2ps}) and (\ref{eq:error2nps}) that the ratio $\beta _{\mathrm{ps}}(g) /\beta _{\mathrm{nps}}(g)$ becomes large.

To summarize this section, if $|\Aw| \ge 1$, the WVA has the advantage for the reduction of the type-2 error while keeping the type-1 error fixed, regardless of the coupling constant $g$, the initial fluctuation $\sigma$, and the critical point $c$.
Namely, the weak measurement more reduces the possibility of missing the presence of the interaction with
the false alarm rate fixed than the measurement without postselection,
if the weak value is outside of the normal range of the eigenvalues.
We emphasize that our result gives a different physical intuition of the WVA that the weak value can be figure of merit for the detection power.


\subsection{A proof that our test is UMPU}
\label{Sec. 3B}

We prove that our decision function (\ref{eq:SNdecision}) is the UMPU test for the probability distributions given by the each measurement in accordance with Lemma giving the tests (\ref{eq:UMPUstatistics}) and (\ref{eq:extendlikehoodfunction}).
We can easily see that the functions (\ref{eq:fps}) and (\ref{eq:fnps}) satisfy \eq{eq:UMPUconditionsmooth}. 
We have shown \eq{eq:1sterror} which indicates that Eq. (\ref{eq:UMPUlemmacondition}) can be satisfied by appropriately choosing the critical point $c$ in each measurement.
We can also show that Eq. (\ref{eq:UMPUlemmacondition2}) is satisfied as
\begin{align}
&\partial_{g} \beta_{\mathrm{ps}}(g) \big| _{g=0} 
=\partial _{g} (1- \cPr{\mE_{2,\mathrm{ps}}} )\big| _{g=0} \notag \\
&=
\frac{-1
}{\big[ 1+ | \Aw |^2
+ ( 1- | \Aw | ^2\big) e^{-\frac{g^2}{2 \sigma ^2}}
 \big]} \notag \\
&~\left. \times
\left[
\begin{array}{l} 
 \frac{-1}{\sqrt{2\pi\sigma^2}}( 1+ | \Aw | ^2) 
\big( e^{-\frac{(c\sigma-g)^2}{2\sigma^2}}-e^{-\frac{(c\sigma+g)^2}{2\sigma^2}}\big)
\\
+\frac{g}{\sigma ^{2}}( 1- | \Aw | ^2) e^{-\frac{g^2}{2 \sigma ^2}}
\erf\big[\frac{c}{\sqrt{2}}\big]
\\
+\frac{g}{\sigma ^{2}}( 1- | \Aw | ^2) \cPr{ \mE_{2,\mathrm{ps}}} e^{-\frac{g^{2}}{2\sigma ^{2} }} 
\end{array}
\right] 
\right| _{g=0} \notag\\
&=0,
\end{align} 
and
\begin{align}
\partial_{g} \beta_{\mathrm{nps}}(g) \big| _{g=0} &=\partial _{g} (1- \cPr{\mE_{2,\mathrm{nps}}} )\big| _{g=0} \notag \\
&=\frac{1}{\sqrt{2\pi\sigma^2}}\big( e^{-\frac{(c\sigma-g)^2}{2\sigma^2}}-e^{-\frac{(c\sigma+g)^2}{2\sigma^2}}\big)\big| _{g=0} \notag \\
&=0.
\end{align} 
Here, we consider the case of the weak measurement.
Equation (\ref{eq:extendlikehoodfunction}) with $f_{\mathrm{ps}} (x |g) $ becomes
\begin{align}
\frac{\mathcal{F}_{\mathrm{ps}} (x)}{f_{\mathrm{ps}}(x|0)}
&=\mathcal{G} _{\mathrm{ps}}(x)-c_{1,\mathrm{ps}} -c_{2,\mathrm{ps}}\frac{\Re \Aw}{\sigma^2}x,
\end{align} 
where
\begin{align}
\mathcal{G}_{\mathrm{ps}} (x):= & \frac{e^{-\frac{g^{2}}{2 \sigma ^2}}}{2 \big[ 1 + | \Aw |^2 + (1 - | \Aw |^2 )  e^{-\frac{g^2}{2\sigma ^2} }\big]} \notag \\
&\times  \left\{ 
\begin{array}{l}
    ( 1 + | \Aw |^2 +2 \Re \Aw )  e ^{\frac{xg}{\sigma^2}} \\
   +( 1 + | \Aw |^2 -2 \Re \Aw)  e ^{-\frac{xg}{\sigma^2}} \\
   +2 ( 1 - | \Aw |^2 )
 \end{array} 
\right\}.
\end{align}
Then, we can find $c_{1,\mathrm{ps}}$ and $c_{2,\mathrm{ps}}$ which give $x=\pm c\sigma$ for the solutions of $\mathcal{F}_{\mathrm{ps}} (x)/f_{\mathrm{ps}}(x|0)=0$ as follows:
\begin{align}
c_{1,\mathrm{ps}}&= \frac{\mathcal{G} _{\mathrm{ps}}(c\sigma)+\mathcal{G} _{\mathrm{ps}}(-c\sigma)}{2}\notag \\
&=\frac{e^{-\frac{g^{2}}{2 \sigma ^2}}
}{1+ | \Aw |^2+ ( 1- | \Aw | ^2) e^{-\frac{g^2}{2 \sigma ^2}}} \notag \\
&~~\times
\left[
\begin{array}{l} 
2(1 + |\Aw |^2 )  \cosh \left[ \frac{cg}{\sigma}\right]\\
+ (1 - |\Aw |^2)
\end{array}
\right]  , \\
c_{2,\mathrm{ps}}&=\sigma \frac{\mathcal{G} _{\mathrm{ps}}(c\sigma) -\mathcal{G} _{\mathrm{ps}}(-c\sigma)}{2\Re \Aw c} \notag \\
&=\frac{ 2\sigma e^{-\frac{g^{2}}{2 \sigma ^2}} \sinh \left[ \frac{cg}{\sigma^2} \right] }{c\big[ 1 + | \Aw |^2 + (1 - | \Aw |^2 )  e^{-\frac{g^2}{2\sigma ^2} }\big]  },
\end{align}
for $\Re \Aw\neq0$.
Because the $\mathcal{G} _{\mathrm{ps}}(x)$ is a convex function, we can interpret
the UMPU test given by the Lemma as the proposed decision function (\ref{eq:SNdecision}).
Therefore, the test (\ref{eq:SNdecision}) with $f_{\mathrm{ps}} (x |g) $ is UMPU.
Even if $\Re \Aw=0$, the discussion remains valid.

A similar discussion holds in the case of the measurement without postselection.
In this case, the distribution function is $f_{\mathrm{nps}}(x| g)$ and \eq{eq:extendlikehoodfunction} becomes
\begin{align}
\frac{\mathcal{F}_{\mathrm{nps}} (x)}{f_{\mathrm{nps}}(x|0)}
&=\mathcal{G} _{\mathrm{nps}}(x)-c_{1,\mathrm{nps}} -c_{2,\mathrm{nps}}\frac{|\langle + | i \rangle |^2-|\langle - | i \rangle |^2}{\sigma^2}x,
\end{align} 
where
\begin{align}
\mathcal{G}_{\mathrm{nps}} (x):= e^{-\frac{g^2}{2\sigma^2}}\big(|\langle + | i \rangle |^2 ~e ^{ \frac{xg}{ \sigma ^2} }+|\langle - | i \rangle |^2 ~e ^{- \frac{xg}{\sigma ^2} }\big).
\end{align}
Then, we can obtain $c_{1,\mathrm{nps}}$ and $c_{2,\mathrm{nps}}$ which give $x=\pm c\sigma$ as the solutions of $\mathcal{F}_{\mathrm{nps}} (x)/f_{\mathrm{nps}}(x|0)=0$ as follows:
\begin{align}
c_{1,\mathrm{nps}}&= \frac{\mathcal{G} _{\mathrm{nps}}(c\sigma)+\mathcal{G} _{\mathrm{nps}}(-c\sigma)}{2}=e^{-\frac{g^2}{2\sigma^2}}\cosh\left[\frac{cg }{\sigma}\right],\\
c_{2,\mathrm{nps}}&=\sigma\frac{\mathcal{G} _{\mathrm{nps}}(c\sigma) -\mathcal{G} _{\mathrm{nps}}(-c\sigma)}{2\left(|\langle + | i \rangle |^2-|\langle - | i \rangle |^2\right) c}=\frac{\sigma e^{-\frac{g^2}{2\sigma^2}}}{c } \sinh \left[ \frac{cg }{\sigma}\right].
\end{align}
Here, we have taken $|\langle + | i \rangle |^2\neq|\langle - | i \rangle |^2$ to obtain the $c_{2,\mathrm{nps}}$.
Even if $|\langle + | i \rangle |^2=\bigl|\langle- | i \rangle \bigr|^2$, the discussion goes through.
The \small $\mathcal{G} _{\mathrm{nps}}(x)$ is also a convex function.
Therefore, we have found that the decision function (\ref{eq:SNdecision}) is the UMPU test in the both measurement cases.


\subsection{Obtaining UMP test for detection problem under certain conditions}
\label{Sec. 3C}

We show that our decision function (\ref{eq:SNdecision}) gives a UMP test in the case $\Re \Aw=0$ for the weak measurement and the case $\bigl| \langle + | i \rangle \bigr|^{2}=\bigl| \langle - | i \rangle \bigr| ^{2}$ for the measurement without postselection.
In these cases, the distribution functions (\ref{eq:fps}) and ~(\ref{eq:fnps}) become even, and the sign of $g$ becomes indistinct. Then, we can practically assume the sign of $g$ is positive without losing generality.
According to the following theorem, we can obtain the UMP test for composite hypotheses such as $H_0:~ \theta \leq\theta _0$ and $H_1:~ \theta > \theta _0 $.

{\it Theorem.}If the likelihood-ratio becomes a monotonically increasing function of the statistics $T(x)$ which is composed of the sample data $x$, 
the following test becomes the UMP for hypotheses: $H_0:~ \theta \leq\theta _0$ and $H_1:~ \theta > \theta _0 $~\cite{Rao1973, Lehmann1959};
\begin{align}
\label{eq:UMPstatistics}
d (x) 
=\left\{
\begin{array}{l}
0 ~~~~\text{if} ~  T(x) < c,  \\
r ~~~~\text{if} ~  T(x) = c, \\
1 ~~~~\text{if} ~  T(x)  > c.
\end{array}
\right.
\end{align}

Because the Theorem is applicable to a one-side test such as $H_0:~ g=0 $ and $H_1:~ g > 0 $, we can show that the test (\ref{eq:SNdecision}) becomes UMP for the each measurement under the certain conditions which make the distribution functions after the measurement even.

We can calculate the likelihood ratios from Eqs. (\ref{eq:fps}) and (\ref{eq:fnps}) in the each case as
\begin{gather}
\label{eq:likefliyps}
\frac{f _ {\mathrm{ps}} (x | g  ) }{f _ {\mathrm{ps}} (x | g =0) }=
\frac{ 1- (\Im \Aw)^2  +[1+ (\Im \Aw)^2 ] \cosh \left[ \frac{xg }{\sigma ^2}\right]
}{
1- (\Im \Aw)^2 +\big[1 +(\Im \Aw)^2\big]  e^{\frac{g^2}{2\sigma ^2} }
},  \\
\label{eq:likefliynps} 
\frac{f _ {\mathrm{nps}} (x | g  ) }{f _ {\mathrm{nps}} (x | g =0) }=
 e^{-\frac{g^2}{2 \sigma ^2}}\cosh \left[ \frac{xg}{  \sigma ^2} \right],
\end{gather}
respectively.
To obtain them, we have used $\Re \Aw=0$ and $\bigl| \langle + | i \rangle \bigr|^{2}=\bigl| \langle - | i \rangle \bigr| ^{2}$.
Since $\cosh$ is an even function, 
\begin{align}
\cosh\left[\frac{gx}{\sigma^2}\right]=\cosh \left[\frac{g}{\sigma} \frac{|x|}{\sigma} \right].
\end{align}
For the ratios (\ref{eq:likefliyps}) and (\ref{eq:likefliynps}) to be a UMP test, we demand Eqs. (\ref{eq:likefliyps}) and (\ref{eq:likefliynps}) are functions of statistics $T(x)$ which are independent of the unknown parameter $g$.
Then, we can set the statistics $T(x)=|x|/\sigma$ not to contain the unknown parameter $g$ so that the likelihood ratios (\ref{eq:likefliyps}) and (\ref{eq:likefliynps}) become monotonically increasing functions.
Thus, we find the statistics $T(x)=|x|/\sigma$ and the Theorem that our test (\ref{eq:SNdecision}) is UMP in the case $\Re \Aw=0$ for the weak measurement and the case $\bigl| \langle + | i \rangle \bigr|^{2}=\bigl| \langle - | i \rangle \bigr| ^{2}$ for the measurement without postselection.


\section{TESTING IN THE CASE WITH AN ADDITIVE WHITE GAUSSIAN NOISE}
\label{Sec. 4}

In this section, we remark on the noise tolerance of the hypothesis test proposed in the previous section.
The probability distribution of the experimental result $x$ is ideally given by the distribution functions (\ref{eq:fps}) or (\ref{eq:fnps}).
However, there is always noise.
We assume that the noise $y$ is added to $x$ by passing through the device circuit with the Gaussian probability $e^{-y^2/2s^2}/\sqrt{2 \pi s^2}$ with an arbitrary fluctuation $s$.
This noise model is widely known as an additive white Gaussian noise.
This is seen in the thermal noise generated in an electrical conductor or the shot noise in an electronic circuit.
This Gaussian jitter noise in the estimation accuracy of the WVA was discussed in Ref.~\cite{Knee2014_2}.
We have the distribution of $z=x+y$ from the moment-generating functions of $x$ and $y$ distributions, and we show its derivation in Appendix \ref{App. B}.

From the distribution of $z$ we have the probabilities of the two types of errors.
The probability of the type-1 error with the postselection and the one without postselection are the same as,
\begin{align}
\label{eq:1sterror_n}
\cPr{\mE_{1,\mathrm{ps}}}=\cPr{\mE_{1,\mathrm{nps}}} =1- \erf \left[ \frac{c\sigma}{ 2\sqrt{\sigma ^2 +s^2} }\right] .
\end{align} 
The probability of the type-2 error with the postselection is
\begin{align}
\label{eq:error2ps_n}
&\cPr{\mE_{2,\mathrm{ps}}} =1-\beta_{\mathrm{ps}}(g) \notag \\
&=\frac{1}{2 \big[ 1+ | \Aw |^2
+ ( 1- | \Aw | ^2) e^{-\frac{g^2}{2 \sigma ^2}}
 \big] } 
\notag   \\
&
\times
\left[
\begin{array}{l}  
( 1+ | \Aw | ^2 )
\big( 
   \erf \big[\frac{c\sigma-g}{\sqrt{2(\sigma^2 +s^2)}} \big] + 
   \erf \big[\frac{c\sigma+g}{\sqrt{2(\sigma^2 +s^2)}} \big]   
 \big) \\
+2( 1- | \Aw | ^2) e^{-\frac{g^2}{2 \sigma ^2}}
  \erf \big[\frac{c\sigma}{2\sqrt{(\sigma^2 +s^2)}} \big]
\end{array}
    \right] , 
\end{align}
and the one without postselection is
\begin{align}
\label{eq:error2nps_n}
&\cPr{\mE_{2,\mathrm{nps}}} =1-\beta_{\mathrm{nps}}(g) \notag \\
&=\frac{1}{2} 
\left( 
\erf \left[ \frac{c\sigma-g}{\sqrt{2(\sigma^2 +s^2)}} \right] 
+ \erf \left[\frac{c\sigma+g}{\sqrt{2(\sigma^2 +s^2)}} \right]  
 \right) . 
\end{align} 

These probabilities are almost the same as Eqs. (\ref{eq:error2ps}) and (\ref{eq:error2nps}), respectively except the denominator of the argument of the error functions.
Comparing these probabilities and the detection powers as in Sec. \ref{Sec. 3A}, we can extend the conclusion that the inequalities (\ref{eq:ineq}) and (\ref{eq:ineq_power}) hold $|\Aw|\geq1$ with an additive white Gaussian noise.
Therefore, we conclude that our testing is robust against the unknown fluctuation.


\section{SUMMARY AND DISCUSSION}
\label{Sec. 5}

In this paper, we have studied the capability of the WVA to detect whether the interaction is present or not in an indirect quantum measurement scheme with the statistical hypothesis testing.
We conclude that the merit of the WVA is the increase of the detection power, which agrees with the previous intuition suggested by Aharanov, Albert, and Vaidman in Ref.~\cite{Aharonov1988}.
Precisely, the WVA reduces the possibility to miss the presence of the interaction with a fixed false alarm rate than the ordinary measurement, when the absolute value of the weak value is greater than the eigenvalues.
We have also shown that our hypothesis testing has the robustness against the additive white Gaussian noise.
Our discussion holds under the assumption that the measured system is the two-state system and that the initial wave function of the measuring probe is Gaussian.

We have proposed the UMPU test for the interaction detection problem, which should be treated as the two-side test.
Our decision function is provided from the intuition that there will be an interaction if the measurement result is outside the initial fluctuation of the probe distribution.
We remark that the proposed test is regarded as a UMP test in the specific case such that the detection problem essentially behaves as a one-side test.

The statistical reliability in the hypothesis testing is given by the probabilities of errors, not by the number of data as explained in Sec. \ref{Sec. 2B}.
Thus, our result holds even for a small number of measurement results.
We note that our result does not conflict with that of the estimation theory in Refs.~\cite{Knee2013, Tanaka2013, Ferrie2014} which needs a large number of data for accurate determination of the parameter.
Generally speaking, we can say about the parameter in more detail by the estimation than by the hypothesis testing.
If data are large, both measurements, the measurement without postselection (including the strong measurement) and the weak measurement, work well for the parameter estimation and the hypothesis testing with an appropriate decision function.
For small data, however, the method of the parameter estimation is not generally reliable for the both measurements, while the weak measurement does a better job than the measurement without postselection for the hypothesis testing as we have shown in Sec. \ref{Sec. 3A}.

At this stage, we need to discuss the case that we cannot obtain any measurement data due to complete failure of the postselection.
To cope with such a case, we consider a makeshift decision function as an attempt to discuss in Appendix \ref{App. C}.
There, we have found that the optimal condition for reducing the type-2 error with the type-1 error under a certain significance level is that the preselected state is the eigenstate of the measured observable and the postselection is not necessary.
However, this discussion has defects on the treatment of the failure of the postselection.
There is no reasonable ground that we regard the failure of the postselection as the absence of interaction, because the postselection can fail whether the interaction is present or absent.
When there is a case that we cannot obtain any data, the problem of the null result arises.
Even though we know how to treat the null result in the projective measurement~\cite{Elitzur1993,Kwiat1995}, that of the null result in the weak measurement has not been developed yet.
This remains an open problem.

Generally speaking, a UMP test and a UMPU test do not always provide an optimal solution 
and it is difficult to optimize the statistical hypothesis testing~\cite{Berger1985}.

\section*{ACKNOWLEDGMENTS}

We appreciate valuable comments from statistical viewpoint by Professor F. Tanaka.
We thank Professor A. Hosoya for reading of manuscript.
Y. S. is supported by JSPS (Grant No. 25008633).

\appendix


\section{PROOF OF \\THE INEQUALITY (\ref{eq:ineq})}
\label{App. A}

Here, we prove that the inequality (\ref{eq:ineq}) as $\cPr{\mathcal{E}_{2,\mathrm{ps}}} \leq \cPr{\mathcal{E}_{2,\mathrm{nps}} }$ when $| \Aw |\geq1$ by looking at the right hand side of \eq{eq:E2ratio}.
More precisely, we show
\begin{align}
\label{eq:ervs1}
\frac{ 2\erf \left[\frac{c}{\sqrt{2}}\right] }
      {\erf \left[\frac{c\sigma-g}{\sqrt{2\sigma^2}} \right] + 
         \erf \left[\frac{c\sigma+g}{\sqrt{2\sigma^2}} \right]}>1.
\end{align}
It is enough to show (\ref{eq:ervs1}) only for the case $g>0$ because the symmetry of the left hand side of the inequality (\ref{eq:ervs1}) under the exchange $g\leftrightarrow-g$.
For the case $0<g\leq c\sigma$, $ e^{-(t-g/\sqrt{2\sigma^2})^{2}} > e^{-t^2}$ holds when $t \ge  c/\sqrt{2}$. Then we have
\begin{align}
\label{eq:cal1}
& \int  _{\frac{c}{\sqrt{2}}}  ^{\frac{c\sigma+g}{\sqrt{2\sigma^2}}}  e^{-(t-\frac{g}{\sqrt{2 \sigma^2}})^{2}}dt 
> 
\int _{\frac{c}{\sqrt{2}}}  ^{\frac{c\sigma+g}{\sqrt{2\sigma^2}}} e^{-t^{2}} dt \notag \\
&\Leftrightarrow 
\int _{\frac{c\sigma-g}{\sqrt{2\sigma^2}}} ^{\frac{c}{\sqrt{2}}} e^{-t^{2}}dt 
>
\int _{\frac{c}{\sqrt{2}}}  ^{\frac{c\sigma+g}{\sqrt{2\sigma^2}}} e^{-t^{2}} dt \notag\\
&\Leftrightarrow 
\erf \left[\frac{c}{\sqrt{2}}\right] -\erf \left[\frac{c\sigma- g }{\sqrt{2\sigma^2}} \right]
>
 \erf \left[\frac{c\sigma+ g}{\sqrt{2\sigma^2}} \right]-\erf \left[\frac{c}{\sqrt{2}}\right] .
\end{align}
Thus, (\ref{eq:ervs1}) is shown for $0<g\leq c\sigma$.
Next we consider the case $g>c\sigma$. 
Because the inequity $ e^{-(t-c/\sqrt{2})^{2}} > e^{-t^2}$ holds for  $t \ge c/\sqrt{2}$, we obtain
\begin{align}
\label{eq:cal2}
&\int  _{\frac{c}{\sqrt{2}}}  ^{\sqrt{2}c}  e^{-(t-\frac{c}{\sqrt{2}})^{2}}dt 
> 
\int  _{\frac{c}{\sqrt{2}}}  ^{\sqrt{2}c}  e^{-t^{2}} dt \notag \\
&\Leftrightarrow 
\int  _{0}  ^{\frac{c}{\sqrt{2}}}  e^{-t^2}dt 
>
\int  _{\frac{c}{\sqrt{2}}}  ^{\sqrt{2}c}  e^{-t^{2}} dt \notag \\
&\Leftrightarrow 
\erf \left[\frac{c}{\sqrt{2}} \right]
>
\erf \left[\sqrt{2}c\right]-\erf \left[\frac{c}{\sqrt{2}} \right].
 \end{align}
Also we have $e^{-(t-\sqrt{2}c)^{2}} > e^{-t^2}$ for $t \ge  \sqrt{2}c$, and
\begin{align}
\label{eq:cal3}
 &\int  _{\sqrt{2}c}  ^{\frac{c\sigma+g}{\sqrt{2\sigma^2}}}  e^{-\left(t-\sqrt{2}c\right)^{2}} dt 
>
\int  _{\sqrt{2}c}  ^{\frac{c\sigma+g}{\sqrt{2\sigma^2}}}  e^{-t^{2}} dt \notag \\
&\Leftrightarrow 
\int  _{0}  ^{-\frac{c\sigma-g}{\sqrt{2\sigma^2}}}  e^{-t^{2}} dt 
>
\int  _{\sqrt{2}c}  ^{\frac{c\sigma+g}{\sqrt{2\sigma^2}}}  e^{-t^{2}} dt \notag \\
&\Leftrightarrow
-\erf\left[ \frac{c\sigma-g}{\sqrt{2\sigma^2}}\right]
>
\erf\left[\frac{c\sigma+g}{\sqrt{2\sigma^2}}\right]-\erf[\sqrt{2}c].
\end{align}
Adding (\ref{eq:cal2}) to (\ref{eq:cal3}), we get 
\begin{align}
\label{eq:cal4}
\erf \left[\frac{c}{\sqrt{2}}\right] -\erf \left[\frac{c\sigma- g }{\sqrt{2\sigma^2}} \right]
>
\erf \left[\frac{c\sigma+ g}{\sqrt{2\sigma^2}} \right]-\erf \left[\frac{c}{\sqrt{2}}\right] .
\end{align}
We have shown (\ref{eq:cal1}) for $0<g\leq c\sigma$ and (\ref{eq:cal4}) for $g>c\sigma$.
Putting them together, we have (\ref{eq:ervs1}) for $g>0$, and therefore $\cPr{\mathcal{E}_{2, \mathrm{ps}}} \leq \cPr{\mathcal{E}_{2,\mathrm{nps}} }$ when $| \Aw |\geq1$.


\section{DERIVATION OF THE $z$ DISTRIBUTION FUNCTION IN SEC. \ref{Sec. 4}}
\label{App. B}

A moment-generating function determines the distribution function of a random variable, and we can derive the distribution of $z$ from its moment-generating function.
Since the random variables $x$ and $y$ are independent, the moment-generating function of $z=x+y$ satisfies
\begin{align}
\E [ e ^{ \xi z } ]
= \E [ e ^{ \xi x } ] \E [ e ^{ \xi y } ],
\end{align}
where the $\E$ means the expectation value.
It is known that the moment-generating function of the Gaussian distribution $N(y) := e^{-y^2/(2s^2) }/\sqrt{2 \pi s^2}$ is,
\begin{align}
\E [ e ^{ \xi y } ]
= \int e ^{ \xi y } N(y) dy
= e ^{\frac{s^2 }{2} \xi^2 } .
\end{align}
Then, the moment-generating function of the $x$ distribution with postselection is
\begin{align}
\E _{f_{\mathrm{ps}}} &[ e ^{ \xi x } ]
= \int e ^{ \xi x } f_{\mathrm{ps}}(x|g ) dx \notag \\
&= 
\frac{e^{\frac{\sigma ^2  }{2} \xi ^2} }{2\big[1 + | \Aw |^2 + (1 - | \Aw |^2)  e^{-\frac{g^2}{2\sigma ^2} }\big]} \notag 
\\&  \times 
\left[
 \begin{array}{l}
( 1 + | \Aw |^2 +2 \Re \Aw )  e ^{g \xi }\\
+( 1 + | \Aw |^2 -2 \Re \Aw)  e ^{-g \xi} \\
+2 ( 1 - | \Aw |^2 )  e ^{-\frac{g^2}{2 \sigma ^2}} 
 \end{array} 
\right] , 
\end{align}
and
\begin{align}
\E _{f_{\mathrm{ps}}} & [ e ^{ \xi z } ]  
= \E _{f_{\mathrm{ps}}} [ e ^{ \xi x } ] \E [ e ^{ \xi y } ] \notag \\
&= 
\frac{e^{\frac{\sigma ^2 +s^2 }{2} \xi ^2} }{2\big[1 + | \Aw |^2 + (1 - | \Aw |^2 )  e^{-\frac{g^2}{2\sigma ^2} }\big]} 
\notag \\&  \times 
\left[
 \begin{array}{l}
( 1 + | \Aw |^2 +2 \Re \Aw )  e ^{g \xi }\\
+( 1 + | \Aw |^2 -2 \Re \Aw)  e ^{-g \xi} \\
+2 ( 1 - | \Aw |^2 )   ~e ^{-\frac{g^2}{2 \sigma ^2}} 
 \end{array} 
\right].
\end{align}
Because of the linearity of the expectation value, the distribution of $z$ is given by
\begin{align}
\label{eq:AWGpsDF}
& f_{\mathrm{ps}}(z|g) \notag \\
&=\frac{1}{2\sqrt{2\pi( \sigma ^2 +s^2 ) } }
\frac{1}{1 + | \Aw |^2 + (1 - | \Aw |^2 )  e^{-\frac{g^2}{2\sigma ^2} }} 
 \notag \\&  \times 
\left[
 \begin{array}{l}
( 1 + | \Aw |^2 +2 \Re \Aw )  ~e ^{-\frac{(x-g)^2}{2 ( \sigma ^2 +s^2 ) }} \\
+( 1 + | \Aw |^2 -2 \Re \Aw)  ~e ^{-\frac{(x+g)^2}{2 ( \sigma ^2 +s^2 ) }} \\
+2 ( 1 - | \Aw |^2 )   ~e ^{-\frac{g^2}{2 \sigma ^2}}
e ^{-\frac{x ^2}{2( \sigma ^2 +s^2 ) }} 
 \end{array} 
 \right].  
\end{align}
Similarly, we have the moment-generating function of the $x$ distribution without postselection
\begin{align}
\E _{f_{\mathrm{nps}}} [ e ^{\xi x } ]   
= & e^{\frac{\sigma ^2 }{2} \xi ^2} 
\big\{ |\langle + | i \rangle |^2 ~e ^{g \xi }
+|\langle - | i \rangle |^2 ~e ^{-g \xi}
\big\},
\end{align}
and the distribution of $z$,
\begin{align}
\label{eq:AWGnpsDF}
& f_{\mathrm{nps}} (z|g) 
= \frac{
|\langle + | i \rangle |^2 ~e ^{- \frac{(z-g)^2}{2 ( \sigma^2 +s^2 ) } }
+|\langle - | i \rangle |^2 ~e ^{- \frac{(z+g)^2}{2( \sigma^2 +s^2 ) } }
}{\sqrt{2\pi( \sigma^2 +s^2 ) } } .
\end{align}
We have obtained (\ref{eq:AWGpsDF}) and (\ref{eq:AWGnpsDF}), from which we can calculate the probabilities of the type-1 error \eq{eq:1sterror_n} and those of the type-2 error Eqs. (\ref{eq:error2ps_n}) and (\ref{eq:error2nps_n}) in Sec. \ref{Sec. 4}.

The distribution functions (\ref{eq:AWGpsDF}) and (\ref{eq:AWGnpsDF}) can be derived by convolution, which is known to give a Gaussian channel.
We can calculate the distribution function with postselection
\begin{align}
\label{eq:convolutionps}
\int f_\mathrm{ps} (x|g)N(z-x) dx
\end{align}
by using for example
\begin{align}
& \int \frac{e ^{-\frac{(x-g )^2}{2 \sigma ^2}} }{\sqrt{2 \pi \sigma ^2} } \frac{e ^{-\frac{(z-x )^2}{2 s^2}} }{\sqrt{2 \pi s^2} } dx \notag \\
&=\frac{1}{\sqrt{(2 \pi \sigma^2 )(2 \pi s^2 )} }\int e^{- \frac{ (\sigma ^2 +s^2) x^2 -2 (\sigma^2 z +s^2g ) x + s^2 g^2 +\sigma ^2 z^2 }{2\sigma ^2 s^2}} dx \notag \\
& =\frac{e^{ \frac{ (s^2g +\sigma ^2 z )^2 }{2 \sigma ^2 s^2 (\sigma ^2 +s^2) }- \frac{ s^2g^2 +\sigma ^2 z^2 }{2 \sigma ^2 s^2} }}{\sqrt{(2 \pi \sigma^2 )(2 \pi s^2 )} }\int e^ { -\frac{ (\sigma ^2 +s^2) }{2 \sigma ^2 s^2 }\left( x- \frac{\sigma ^2 z +s^2g}{\sigma ^2+s^2} \right)^2} dx \notag \\
&=\sqrt{\frac{2 \pi \sigma^2 s^2 }{(2 \pi \sigma^2 )(2 \pi s^2 )(\sigma ^2+s^2 )} }e^{ - \frac{ \sigma ^2 s^2( z^2 - 2 gz +g^2) }{2 \sigma ^2 s^2 (\sigma ^2 +s^2)  }} \notag \\
&=\frac{1}{\sqrt{2 \pi (\sigma ^2+s^2 )} } e^{ - \frac{ (z-g )^2 }{2 (\sigma ^2+ s^2) }}.
\end{align}
We immediately see that Eq. (\ref{eq:convolutionps}) equals to Eq (\ref{eq:AWGpsDF}) and that a Gaussian channel and an additive white Gaussian noise are identical.


\section{ EVALUATION OF THE TEST INCLUDING THE LOSS BY POSTSELECTION USING THE LAGRANGE MULTIPLIER METHOD}
\label{App. C}

In Sec. \ref{Sec. 3}, we have considered the hypothesis testing if it was able to acquire the data at least once.
Here, we consider the risk that the data cannot be obtained by failure of the postselection and discuss it by including it in the cost function~\cite{cost function}.
To obtain the optimized process, we minimize this function by the Lagrange multiplier method.

In order to treat the data loss by failure of the postselection taking the test (\ref{eq:SNdecision}) into account,  
we propose a revised test by the following decision function:
\begin{align}
\label{eq:doubletest}
d(x) := \left\{ 
\begin{array}{l}
1 ~~~~\text{if}~ (f~\text{and} ~ |x| > c_f  \sigma)~\text{or}~(\bar{f}~\text{and} ~ |x| > c_{\bar{f}} \sigma),  \\
0 ~~~~\text{if}~ (f~\text{and} ~|x| < c_f  \sigma)~\text{or}~( \bar{f}~\text{and} ~ |x| < c_{\bar{f}} \sigma),   \\
r  ~~~~\text{otherwise}.
\end{array}
\right.
\end{align} 
We denote $f$ and $\bar{f}$ as success and failure of the postselection, respectively.
In this test, we use the postselection result and the measurement result of the probe as statistics.
The critical points $c_f$ and $c_{\bar{f}}$ differ depending on the result of postselection, success or failure.

Here, we calculate the probabilities of the two types of error.
The probability of the type-1 error is
\begin{align}
\label{eq:double1st}
\cPr{\mE_1}=&\Pr [d=1 | g=0 ] \notag \\
=&\Pr [ f , |x| > c_f  \sigma | g=0] +\Pr [ \bar{f} , |x| > c_{\bar{f}}  \sigma | g=0] \notag \\
=&1 -\left(  \erf  \left[\frac{c_f}{\sqrt{2} }\right] | \langle f | i\rangle| ^2 
       + \erf  \left[\frac{c_{\bar{f}}}{\sqrt{2} }\right] | \langle \bar{f} | i\rangle | ^2 \right),
\end{align}
and the probability of the type-2 error is
\begin{align}
\label{eq:double2nd}
&\cPr{\mE_2}\notag \\
&~=\Pr [d=0 | g\neq 0 ] \notag \\
&~= \Pr [f ,  |x| < c_f  \sigma | g\neq 0] +\Pr [\bar{f} ,  |x| < c_{\bar{f}}  \sigma | g\neq 0] \notag \\
&~=\frac{1}{4} (|\langle f | i \rangle |^2+|\langle f | \hat{A} | i \rangle |^2) 
    \left( \erf  \left[\frac{c_f\sigma-g}{\sqrt{2 \sigma ^2} }\right] + \erf  \left[\frac{c_f\sigma+g}{\sqrt{2 \sigma ^2} }\right] \right)   \notag \\
&~+\frac{1}{4} (|\langle \bar{f} | i \rangle |^2+|\langle \bar{f} | \hat{A} | i \rangle |^2) 
    \left( \erf  \left[\frac{c_{\bar{f}}\sigma-g}{\sqrt{2 \sigma ^2} }\right] + \erf  \left[\frac{c_{\bar{f}}\sigma+g}{\sqrt{2 \sigma ^2} }\right] \right) \notag \\
&~+\frac{1}{2} ( |\langle f | i \rangle |^2-|\langle f | \hat{A} | i \rangle |^2 ) e^{-\frac{g^ 2}{2 \sigma ^2} } \erf \left[ \frac{c_f}{ \sqrt{2} } \right]  \notag \\
&~+\frac{1}{2}( |\langle \bar{f} | i \rangle |^2-|\langle \bar{f} | \hat{A} | i \rangle |^2) e^{-\frac{g^ 2}{2 \sigma ^2} } ~\erf  \left[\frac{c_{\bar{f}}}{\sqrt{2} }\right].
\end{align}

If $c_f =c_{\bar{f}}(=c)$, the probability of the type-1 error becomes
\begin{align}
\cPr{\mE_1}=1 - \erf  \left[\frac{c}{\sqrt{2 } }\right] ,
\end{align}
and the probability of the type-2 error becomes
\begin{align}
\cPr{\mE_2}=\frac{1}{2} \left( \erf  \left[\frac{c\sigma-g}{\sqrt{2 \sigma ^2} }\right] +\erf  \left[\frac{c\sigma+g}{\sqrt{2 \sigma ^2} }\right] \right).
\end{align}
Both are independent of the postselection result.
Then, we can substantially simplify the treatment of the case without postselection when $c_f =c_{\bar{f}}$ by the decision function (\ref{eq:doubletest}).

Let us consider how we express the errors of the weak measurement in the decision function (\ref{eq:doubletest}).
When the postselection fails in the weak measurement experiment, we cannot obtain a measurement result.
So, in such a case, we cannot distinguish whether there is the interaction or not.
But, for convenience, we simply presume that there would not be the interaction.
Since we are interested in the detection of the interaction, the result of no interaction is meaningless.
We can conveniently handle this situation by setting $c_{\bar{f}}=\infty$ in our decision function.
If $c_{\bar{f}}$ were $\infty$, the alternative hypothesis would be always rejected by the test (\ref{eq:doubletest}) when the postselection fails.
Hence, the decision function (\ref{eq:doubletest}) would cover the case with or without postselection including the data loss by failure of the postselection.

Meanwhile, if there is the situation such that we want to detect ``vanishment" of an interaction which usually exists, the treatment of $c_{\bar{f}}$ as stated above is inconsequent.
Here, we note that if the $c_f$ and $c_{\bar{f}}$ take the other value as presented above, we cannot give an obvious interpretation what the experimental situation means.
Thus, it is often difficult to find out the physical meaning of the optimization of the $c_f$ and $c_{\bar{f}}$.
While such problems are remaining, we try out this Lagrange multiplier method.

From here, we calculate the critical points and the initial and final states of the measured system which optimize the test (\ref{eq:doubletest}) by the Lagrange multiplier method.
To optimize the probability of the type-2 error while keeping the probability of the type-1 error at the significance level $\alpha$ which is an arbitrary constant, we set the Lagrangian as
\begin{align}
\mL & (p_1,~p_2,~c_f,~c_{\bar{f}},~ \lambda ) \notag \\
&=\Pr [\mE_2 ]+ \lambda ( \Pr [\mE_1] -\alpha) \notag \\
& = \frac{1}{4} \left[ 
    \begin{array}{l} 
    ( p_1 +p_2 ) \big( \erf  \big[\frac{c_f\sigma-g}{\sqrt{2 \sigma ^2} }\big] + \erf\big[\frac{c_f\sigma+g}{\sqrt{2 \sigma ^2} }\big] \big) \\
    +( 2- p_1 -p_2) \big( \erf  \big[\frac{c_{\bar{f}}\sigma-g}{\sqrt{2 \sigma ^2} }\big] + \erf  \big[\frac{c_{\bar{f}}\sigma+g}{\sqrt{2 \sigma ^2} }\big] \big) \\
    +2 ( p_1 -p_2 )\big( \erf  \big[\frac{c_f}{\sqrt{2} }\big] - \erf  \big[\frac{c_{\bar{f}}}{\sqrt{2}}\big] \big)  e^{-\frac{g^ 2}{2 \sigma ^2} }
    \end{array}
\right]
\notag \\
&~~+ \lambda \left\{ \left[ 
    \begin{array}{l} 
    p_1 \big( 1- \erf  \big[\frac{c_f}{\sqrt{2}}\big] \big) \\
    +(1-p_1) \big( 1- \erf  \big[\frac{c_{\bar{f}}}{\sqrt{2}}\big] \big) 
    \end{array}
\right]
-\alpha \right\}, 
\end{align}
where $\lambda$ is the Lagrange multiplier and the constraint condition comes from the standard strategy of the hypothesis testing as described in Sec. \ref{Sec. 2B}.
To simplify the notations, we denote $p_1 := |\langle f| i \rangle|^2$ and $p_2 := |\langle f | \hat{A} | i \rangle |^2$.
Note that $\hat{A}^2=1$ and $0 < p_1, p_2 < 1$.
Varying the Lagrangian $\mL$ with respect to $\lambda$, the constraint condition reappears as 
\begin{align}
\label{eq:dfdl}
0=\frac{\partial \mL}{\partial \lambda}=\left[ 
   \begin{array}{l} 
    p_1 ( 1- \erf  \big[\frac{c_f}{\sqrt{2} }\big] ) \\
    +(1-p_1) ( 1- \erf  \big[\frac{c_{\bar{f}}}{\sqrt{2}}\big] )
   \end{array}
\right]
-\alpha.
\end{align}
Then, varying the Lagrangian $\mL$ with respect to $p_1$ and $p_2$, we get
\begin{align}
\label{eq:dfdp1} 
0=\frac{\partial \mL}{\partial p_1} 
&= \frac{1}{4} \left[ 
    \begin{array}{l} 
     \erf  \big[\frac{c_f\sigma-g}{\sqrt{2 \sigma ^2}}\big] + \erf \big[\frac{c_f\sigma+g}{\sqrt{2 \sigma ^2}}\big] \\
    -\erf  \big[\frac{c_{\bar{f}}\sigma-g}{\sqrt{2 \sigma ^2}}\big]- \erf  \big[\frac{c_{\bar{f}}\sigma+g}{\sqrt{2 \sigma ^2}}\big]  \\
    +2  \big( \erf  \big[\frac{c_f}{\sqrt{2}}\big] - \erf  \big[\frac{c_{\bar{f}}}{\sqrt{2}}\big] \big)  e^{-\frac{g^ 2}{2 \sigma ^2}}
    \end{array}
\right] \notag \\
&+ \lambda \left( 
- \erf  \left[\frac{c_f}{\sqrt{2}}\right] 
+\erf  \left[\frac{c_{\bar{f}}}{\sqrt{2}}\right] 
\right)
\end{align}
and
\begin{align}
\label{eq:dfdp2}
0=\frac{\partial \mL}{\partial p_2}
 &= \frac{1}{4} \left[ 
    \begin{array}{l} 
     \erf  \big[\frac{c_f\sigma-g}{\sqrt{2 \sigma ^2} }\big] + \erf  \big[\frac{c_f\sigma+g}{\sqrt{2 \sigma ^2} }\big] \\
    -\erf  \big[\frac{c_{\bar{f}}\sigma-g}{\sqrt{2 \sigma ^2} }\big]- \erf  \big[\frac{c_{\bar{f}}\sigma+g}{\sqrt{2 \sigma ^2} }\big] \\
    -2\big( \erf  \big[\frac{c_f}{\sqrt{2}}\big] - \erf  \big[\frac{c_{\bar{f}}}{\sqrt{2}}\big] \big)  e^{-\frac{g^ 2}{2 \sigma ^2} }
    \end{array}
\right].
\end{align}
Form these equations, we have
\begin{align}
\label{eq:dfdp1-dfdp2}
0&=\frac{\partial \mL}{\partial p_1} - \frac{\partial \mL}{\partial p_2} \notag  \\
& =  \left( 
\erf  \left[\frac{c_f}{\sqrt{2} }\right] 
- \erf  \left[\frac{c_{\bar{f}}}{\sqrt{2} }\right] 
\right) \left( 
e^{-\frac{g^ 2}{2 \sigma ^2} }-\lambda
\right).
\end{align}
So, we require either or both of $\lambda= e^{-g^ 2/2 \sigma ^2 }$ and $c_f=c_{\bar{f}} $.
Then, varying $\mL$ with respect to $c_f$ and $c_{\bar{f}}$ gives
\begin{align}
\label{eq:dfdc1}
&0=\frac{\partial \mL}{\partial c_f}  \notag \\
&= \frac{1}{4} \sqrt{\frac{2}{\pi}}\left[  
  \begin{array}{l} 
    ( p_1 + p_2 )  \big(e^{ -\frac{(c_f\sigma-g)^2 }{2 \sigma ^2 }} +  e^{ -\frac{( c_f\sigma+g)^2 }{2 \sigma ^2 }}\big) \\
    +2  \big( p_1 - p_2 -  2\lambda p_1e^{ \frac{g^2 }{2 \sigma ^2 }} \big) e^{ -\frac{c_f^2\sigma^2+g^2 }{2 \sigma ^2 }} 
   \end{array}
\right]
\end{align}
and
\begin{align}
\label{eq:dfdc2}
&0=\frac{\partial \mL}{\partial c_{\bar{f}}} \notag \\
&= \frac{1}{4} \sqrt{\frac{2}{\pi}}\left[  
   \begin{array}{l} 
    ( 2-p_1 -p_2 )  \big(e^{ -\frac{(c_{\bar{f}}\sigma-g)^2 }{2 \sigma ^2 }} +  e^{ -\frac{( c_{\bar{f}}\sigma+g)^2 }{2 \sigma ^2 }}\big) \\
    +2  \big( -p_1 + p_2 -  2\lambda (1-p_1) e^{ \frac{g^2 }{2 \sigma ^2} } \big) e^{ -\frac{c_{\bar{f}}^2\sigma^2+g^2 }{2 \sigma ^2 }} 
   \end{array}
\right] .
\end{align}

Here, we consider the case $c_f=c_{\bar{f}}=c$, where Eqs. (\ref{eq:dfdp1}) and (\ref{eq:dfdp2}) are fulfilled.
The constraint constant (\ref{eq:dfdl}) becomes
\begin{align}
\label{eq:dfdl3}
0=\left( 1- \erf  \left[\frac{c}{\sqrt{2} }\right] \right) -\alpha.
\end{align}
Because the $\alpha$ is a constant, the $c$ is fixed.
Next, from Eqs. (\ref{eq:dfdc1}) and (\ref{eq:dfdc2}) we get
\begin{align}
\label{eq:dfdc13}
0&=\frac{e^{ -\frac{c^2 }{2 }} }{4} \sqrt{\frac{2}{\pi}}
 \left[ 
\begin{array}{l}
  p_1 \big\{ e^{ -\frac{g^2 }{2 \sigma ^2 }}  ( e^{ \frac{cg }{2 \sigma }} +  e^{ -\frac{cg }{2 \sigma }}) ^2-4 \lambda \big\} \\
  +p_2 e^{ -\frac{g^2 }{2 \sigma ^2 }}  \big( e^{ \frac{cg }{2 \sigma}} -  e^{ -\frac{cg }{2 \sigma}}\big) ^2
\end{array}  
\right]
\end{align}
and
\begin{align}
\label{eq:dfdc23}
0&=\frac{e^{ -\frac{c^2 }{2 }} }{4} \sqrt{\frac{2}{\pi}}
 \left[ 
  \begin{array}{l} 
   2 \big\{ e^{ -\frac{g^2 }{2 \sigma ^2 }} ( e^{ \frac{cg }{\sigma }} +  e^{ -\frac{cg }{ \sigma }}) -2\lambda  \big\}\\
 - p_1 \big\{ e^{ -\frac{g^2 }{2 \sigma ^2 }}  ( e^{ \frac{cg }{2 \sigma}} +  e^{ -\frac{cg }{2 \sigma }}) ^2-4 \lambda \big\} \\
  -p_2  e^{ -\frac{g^2 }{2 \sigma ^2 }}  ( e^{ \frac{cg }{2 \sigma}} -  e^{ -\frac{cg }{2 \sigma}}) ^2
\end{array}
 \right],
\end{align}
respectively.
The sum of Eqs. (\ref{eq:dfdc13}) and (\ref{eq:dfdc23}) gives
\begin{align} 
0&= \frac{1}{2} \sqrt{\frac{2}{\pi}}e^{ -\frac{c^2 }{2}}    
  \big\{   e^{ -\frac{g^2 }{2 \sigma ^2 }}  ( e^{ \frac{cg }{\sigma }} +  e^{ -\frac{cg }{ \sigma}}) 
   -2\lambda \big\}.
 \end{align}
Thus, we obtain
\begin{align} 
\label{eq:l3}
\lambda &=\frac{1}{2} e^{ -\frac{g^2 }{2 \sigma ^2 }}  ( e^{ \frac{cg }{\sigma}} +  e^{ -\frac{cg }{ \sigma}}).
\end{align}
Substituting $c_f=c_{\bar{f}}=c$ and \eq{eq:l3} into Eqs. (\ref{eq:dfdc1}) and (\ref{eq:dfdc2}),
we have 
\begin{align}
0=\frac{e^{ -\frac{c^2\sigma^2+g^2 }{2 \sigma ^2 }}}{4} \sqrt{\frac{2}{\pi}} ( p_1 -p_2 )
\big( e^{ \frac{cg }{2 \sigma }} -  e^{ -\frac{cg }{2 \sigma}}\big) ^2.
\end{align}
Then, we can find that $c_f=c_{\bar{f}}=c=0$ or $p_1=p_2$ is needed. 
If $c=0$, we obtain $\alpha=1$ and $\lambda=e^{-g^2/2\sigma^2}$ from Eqs. (\ref{eq:dfdl3}) and (\ref{eq:l3}).
Because the significance level $\alpha$ is not always $1$, the $c=0$ is not consistent.
Then, $c_f=c_{\bar{f}}\neq0$ and $p_{1}=p_{2}$ is a solution.

Next from \eq{eq:dfdp1-dfdp2}, we discuss the case $\lambda=e^{-g^2/2\sigma^2}$.
Because we have already studied the case that $c_f=c_{\bar{f}}$ is satisfied simultaneously, hereafter we assume $c_f\neq c_{\bar{f}}$.
Substituting $\lambda= e^{-g^ 2/2 \sigma ^2}$ into Eqs. (\ref{eq:dfdc1}) and (\ref{eq:dfdc2}), we have
\begin{flalign}
\label{eq:dfdc12}
0=&\frac{1}{4} \sqrt{\frac{2}{\pi}} ( p_1 +p_2 )  \big( e^{ \frac{c_fg }{2 \sigma }} - e^{ -\frac{c_fg }{2 \sigma}}\big) ^2
    e^{ -\frac{c_f^2\sigma^2+g^2 }{2 \sigma ^2 }} ,
\end{flalign}
and
\begin{flalign}
\label{eq:dfdc22}
0=&\frac{1}{4} \sqrt{\frac{2}{\pi}}
    ( 2-p_1 -p_2 )  \big( e^{ \frac{c_{\bar{f}}g }{2 \sigma }} -  e^{ -\frac{c_{\bar{f}}g }{2 \sigma}}\big) ^2
    e^{ -\frac{c_{\bar{f}}^2\sigma^2+g^2 }{2 \sigma ^2 }}.
\end{flalign}
From Eqs. (\ref{eq:dfdc12}) and (\ref{eq:dfdc22}),
we can find that we need either condition as follows: the condition such that $c_f=0$ and $p_1=p_2=1$,
or the condition such that $c_{\bar{f}}=0$ and $p_1=p_2=0$. 
In both cases, the constrain condition (\ref{eq:dfdl}) becomes $0=1-\alpha$.
As stated in above, the $\alpha$ is not necessarily $1$.
So, the condition $\lambda= e^{-g^2/2 \sigma ^2}$ is not proper.

Therefore, we conclude the solution is $c_f=c_{\bar{f}}\neq0$ and $p_1=p_2$.
From $p_1=p_2$, we derive
\begin{align}
 0& =|\langle f| i \rangle |^2-|\langle f | \hat{A} | i \rangle |^2 \notag \\
&=\langle i| (| f \rangle\langle f| -\hat{A}|  f \rangle\langle f|  \hat{A} ) |  i \rangle,  \therefore  \pm | f \rangle = \hat{A} | f \rangle
\end{align}
or
\begin{align}
 0& =\bigl|\langle f| i \rangle \bigr|^2-\bigl|\langle f | \hat{A} | i \rangle \bigr|^2 \notag \\
&=\langle f| (| i \rangle\langle i| -\hat{A}|  i\rangle\langle i|  \hat{A} ) |  f \rangle,  \therefore \pm | i \rangle = \hat{A} | i \rangle.
\end{align}
Then, $p_1=p_2$ means that the preselected state $| i \rangle$ or the postselected state $| f\rangle$ equals to an eigenstate of $\hat{A}$.
As we have noted, the case $c_f=c_{\bar{f}}$ corresponds to the measurement without postselection.
Thus, the result of the postselection has nothing to do with the test. The state of the postselection $| f\rangle$ is not essential.
Consequently, the optimal condition for the test (\ref{eq:doubletest}) is that the preselected state is an eigenstate of the measured observable $\hat{A}$ and that we do not postselect.
We caution the readers that this appendix gives a nothing but a crude trial.
We also note that the Lagrange multiplier method gives the stationary point to the utmost, and they might not be the minimum.



\begin{thebibliography}{99}
\bibitem{randomized test}
In a randomized test, we randomly accept the null hypothesis with the probability $r$.
This test is widely used at the critical point which is not included in neither the accept nor the rejection region.
For the distribution function of continuous variable, the probability for the continuous random variable coincides with
the critical point is zero. This slightly modified test does not change our conclusion.
\bibitem{cost function}
A cost function represents the cost or risk that we have to pay in the process which we want to evaluate. We can choose the cost function as we like.
\end{thebibliography}
\end{document}